\documentclass[a4paper,twocolumn,aps,prd]{revtex4}
\usepackage[latin9]{inputenc}
\usepackage{color}
\usepackage[unicode=true,
 bookmarks=true,bookmarksnumbered=false,bookmarksopen=false,
 breaklinks=false,pdfborder={0 0 0},backref=false,colorlinks=true]
 {hyperref}
\hypersetup{
 linkcolor=blue, citecolor=red}
\usepackage{breakurl}

\makeatletter

\@ifundefined{textcolor}{}
{%
 \definecolor{BLACK}{gray}{0}
 \definecolor{WHITE}{gray}{1}
 \definecolor{RED}{rgb}{1,0,0}
 \definecolor{GREEN}{rgb}{0,1,0}
 \definecolor{BLUE}{rgb}{0,0,1}
 \definecolor{CYAN}{cmyk}{1,0,0,0}
 \definecolor{MAGENTA}{cmyk}{0,1,0,0}
 \definecolor{YELLOW}{cmyk}{0,0,1,0}
}

\usepackage{ulem}
\usepackage{color}
\usepackage{epsfig}

\newcommand{\be}{\begin{equation}}
\newcommand{\ee}{\end{equation}}
\newcommand{\bea}{\begin{eqnarray}}
\newcommand{\eea}{\end{eqnarray}}

\makeatother

\begin{document}

\title{Event-by-event distribution of magnetic field energy over initial fluid energy density in $\sqrt{s_{\rm NN}}$= 200 GeV Au-Au collisions}

\author{ Victor Roy $^{1}$, Shi Pu $^{1}$}

\affiliation{$^{1}$ Institute for Theoretical Physics, Goethe University, 
Max-von-Laue-Str.\ 1, 60438 Frankfurt am Main, Germany}
\begin{abstract}
We estimate the event-by-event~(e-by-e) distribution of the ratio ($\sigma$) of the 
magnetic field energy to the fluid energy density in the transverse plane of
 Au-Au collisions at $\sqrt{s_{\rm NN}}$ = 200 GeV. A Monte-Carlo~(MC) Glauber model is used 
to calculate the $\sigma$  in the transverse plane  for impact parameter b=0, 12 fm
at time $\tau_i\sim$0.5 fm. 
The fluid energy density is obtained by using Gaussian smoothing 
with two different smoothing parameter $\sigma_g$=0.25 , 0.5 fm. 
For $b=0~\rm fm$ collisions $\sigma$ is found to be 
$\ll$ 1 in the central region of the fireball and $\sigma\gtrsim$ 1 at the periphery. 
For b=12 fm collisions $\sigma\gtrsim$ 1. The e-by-e
correlation between $\sigma$ and the fluid energy density ($\varepsilon$) is studied. 
We did not find strong correlation between $\sigma$
and $\varepsilon$ at the centre of the fireball, whereas they are mostly anti-correlated 
at the periphery of the fireball. 
\end{abstract}
\maketitle

\section{Introduction}

The most strongest known magnetic field (B$\sim10^{18} -10^{19}$Gauss) 
in the universe is produced in laboratory  experiments of Au-Au or Pb-Pb 
collisions  in the collider experiments at Relativistic Heavy Ion Collider~(RHIC) and 
at Large Hadron Collider~(LHC). Previous theoretical studies show that the 
intensity of the produced magnetic field rises approximately linearly with the centre of mass energy 
($\sqrt{s_{{\rm NN}}}$) of the colliding nucleons \cite{Bzdak:2011yy,Deng:2012pc}. 
The Lorentz boosted electric fields in such collisions also becomes very 
strong which is  same order of magnitude as magnetic field 
($e\vec{B} \approx e\vec{E} \sim 10 m_{\pi}^2$
for a typical Au-Au collision at top RHIC energy $\sqrt{s_{\rm NN}}$= 200 GeV),
where $m_\pi$ is the pion mass. 
Such intense electric and magnetic fields are strong enough to initiate the
particle production from vacuum via Schwinger mechanism\cite{KEKIntense:2010} . 
Using quantum chromodynamics it was shown in Ref.~\cite{Simonov:2012if}
 that beyond a critical value of magnetic field 
the quark-antiquark state can possibly attain negative mass 
(in the limit of large number of colours).
Thus it is important to know if there is truly an upper limit of magnetic field 
intensity allowed by the quantum chromodynamics when applied in heavy ion collisions.
Or the magnetic field can grow to arbitrary large value with increasing $\sqrt{s_{\rm NN}}$
 as predicted in some earlier studies~\cite{Bzdak:2011yy,Deng:2012pc}. In this work we shall 
calculate the electromagnetic field intensity without considering any such 
constraints, i.e, we assume that the electric and magnetic fields can attain any arbitrary large 
values. 

There are several other interesting recent studies related to the effect 
of ultra-intense magnetic fields in heavy-ion collisions. Here we briefly 
mention a few of them which might be relevant to the present study.
In presence of a strong magnetic field as created in heavy-ion
collisions, a charge current is induced in the Quark Gluon Plasma~(QGP), leading to what
is known as the ``chiral magnetic effect'' (CME)
\cite{Kharzeev:2007jp}.  Within a 3+1  
dimensional anomalous hydrodynamics model a charge dependent hadron azimuthal 
correlations  was found to be sensitive to the CME in Ref.~\cite{Hirono:2014oda}.
Along with CME, it was also predicted theoretically that massless fermions
 with the same charge but different chirality will be separated, yielding what is called
the ``chiral separation effect'' (CSE). A connection between these effects 
and the Berry phase in condensed matter was also pointed out in Ref.~\cite
{Chen:2012ca,Stephanov:2012ki,Son:2012zy}.
In hadronic phase a significant changes in the hadron multiplicity was 
observed in presence of a strong magnetic field  within a statistical hadron 
resonance gas model in Ref.~ \cite{Bhattacharyya:2015pra}. 
There are lot of other important 
relevant work in this new emerging field which we cannot refer here, 
one can see recent reviews on this topic in 
Ref.~ \cite{Kharzeev:2013ffa,Bzdak:2012ia,Kharzeev:2015kna}
for more details.

 The relativistic hydrodynamic models have so far nicely explained the
 experimentally measured anisotropic particle production in the 
 azimuthal directions in heavy ion collisions. 
 The success of hydrodynamics model shows that 
 a locally equilibrated QGP with small ratio of shear viscosity
  to entropy density is formed after the collision within a short time interval 
$\sim$0.2-0.6 fm \cite{Shen:2011zc,Romatschke,
  Heinz, Roy:2012jb, Heinz:2011kt, Niemi:2012ry, Schenke:2011bn}. 
  It is also well known that the final momentum anisotropy in hydrodynamic 
  evolution is very sensitive to the initial (geometry) state of the nuclear collisions.
So far almost all the hydrodynamic models studies have 
neglected any influence of magnetic fields on the initial fluid energy-density or 
on the space-time evolution of  QGP.  But as we know the initial magnetic field 
is quite large, it is important to investigate the relative importance of large 
electro-magnetic field on the usual hydrodynamical evolution of 
QGP. For that one need a full 3+1 dimensional magnetohydrodynamic code to 
numerically simulate the space time evolution of QGP with magnetic fields.  
While one can gain some insight about the relative importance of the magnetic filed on the 
initial energy density of the QGP fluid by estimating the quantity plasma sigma, which is 
the dimensionless ratio of magnetic field energy $\frac{B^2}{2}$ to the fluid 
energy density($\varepsilon$) : $\sigma=\frac{B^2}{2\varepsilon}$. 
 In plasma $\sigma\sim$1 
 indicates that one can no longer neglect the effect of magnetic fields 
 in the plasma evolution (in some situation 
  $\sigma\sim$ 0.01 may also effect the hydrodynamic evolution)
  \cite{Lyutikov:2011vc,Kennel:1983,Roy:2015kma,Spu:2015kma}. 
 In the present study we use MC-Glauber 
 model \cite{Lin:2004en,Miller:2007ri} to calculate e-by-e magnetic fields 
 and fluid energy density in Au-Au collisions at $\sqrt{s_{\rm NN}}=$ 200 GeV 
 and investigate the relative
 importance of the magnetic field on initial fluid energy density. 
  
 As mentioned earlier, the typical magnetic field produced in a 
 mid-central Au-Au collisions at $\sqrt{s_{\rm NN}}=200 \rm GeV$ reaches 
 $\sim10m_{\pi}^{2}$,  which corresponds to 
 field energy density of $\sim 5\,{\rm GeV}/{\rm fm}^{3}$. 
 Hydrodynamical model studies show that the initial
energy density for such cases is $\sim 10\,{\rm GeV}/{\rm fm}^{3}$, thus
implying $\sigma \sim 0.2$ under these conditions. However, the magnetic
field produced at the time of collisions decays very quickly if QGP does not
possess finite electrical conductivity \cite{Gursoy:2014aka, Zakharov:2014dia, 
Tuchin:2013apa}. 
Thus in order to correctly estimate $\sigma$, one need to consider
 the proper temporal evolution of magnetic fields until
the thermalisation time ($\tau_{i}\sim0.5 fm$ for Au-Au collisions at RHIC)
when the hydrodynamic evolution starts.
Since the spatial distribution of fluid energy density as well as the 
electromagnetic fields varies  e-by-e we also calculate $\sigma$ 
accordingly.  The spatial distribution of electric and 
magnetic fields in heavy ion collisions was previously studied in 
Ref.\cite{Zhong:2015dia,Voronyuk:2011jd}.
 
In present work we study
the spatial distribution of $\sigma$ in  $\sqrt{s_{\rm NN}}=200$  GeV Au-Au
collisions for two different impact parameters (b=0, and 12 fm). The temporal evolution of the 
magnetic fields after the collision is taken into account in a simplified manner 
which will be discussed in the next section. We also investigate the correlation 
between $\sigma$ and fluid energy density in the transverse plane. 
The paper is organised as follows: in the next section, we discuss about the formalism.
Our main result and discussion are presented in section III. A summary
is given at the end in section IV.

\section{Formalism and setup}

\begin{figure}
\includegraphics[scale=0.5]{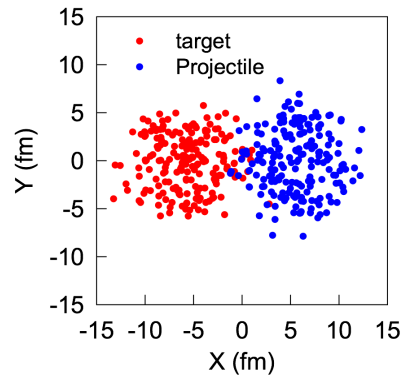}
\caption{Distribution of nucleons inside target and projectile 
nuclei in a typical Au-Au collisions  at $\sqrt{s_{\rm NN}}$
=200 GeV for b=12 fm.}
\label{fig:Npart_b12fm_single}
\end{figure}

We constructed a spatial grid of size 10 fm in each direction (X and Y) 
with the corresponding grid spacing of $\Delta x$ = $\Delta y$ = 0.5 fm
for e-by-e calculation of electromagnetic fields and fluid energy density 
in the plane transverse to the trajectory of the colliding nuclei. The position of colliding 
nucleons are obtained from MC-Glauber model in e-by-e basis. 
The position of nucleons are randomly distributed 
according to the Wood-Saxon nuclear density distribution (as shown
 in Fig~(\ref{fig:Npart_b12fm_single})). 
We adopt the usual convention used in heavy ion collisions
 for describing the geometry of the nuclear collisions
, i.e., the impact parameter vector $(\vec{b})$ of the collision 
is along X axes and the colliding nuclei are symmetrically situated around the 
(0,0) point of the computational grid.  The electric and magnetic fields at point 
$\vec{r} (x,y)$ at time $t$ due to all  charged protons inside  two colliding 
nucleus are calculated from the Lienard-Weichart formula  

\begin{equation}
\vec { E } \left( \vec { r } ,t \right) =\frac { e }{ 4\pi  } \sum _{ i=1 }^{ { N }_{proton} }{ { Z }_{ i }\frac { \vec { { R }_{ i } } -{ R }_{ i }\vec { { v }_{ i } }  }{ { \left( { R }_{ i }-\vec { { R }_{ i } } \cdot \vec { { v }_{ i } }  \right)  }^{ 3 } } \left( 1-{ v }_{ i }^{ 2 } \right),  } 
\label{eq:Efield_LW}
\end{equation}
\begin{equation}
\vec { B } \left( \vec { r } ,t \right) =\frac { e }{ 4\pi  } \sum _{ i=1 }^{ { N }_{proton} }{ { Z }_{ i }\frac { \vec { { v }_{ i } } \times \vec { { R }_{ i } }  }{ { \left( { R }_{ i }-\vec { { R }_{ i } } \cdot \vec { { v }_{ i } }  \right)  }^{ 3 } } \left( 1-{ v }_{ i }^{ 2 } \right)  } .
\label{eq:Bfield_LW}
\end{equation}

Where $\vec{E}$ and $\vec{B}$  are the electric and magnetic field vector respectively, 
e is the charge of an electron,
Z is the number  of proton inside each nucleus, $\vec{R_{i}}=\vec{x} - \vec{x_{i}}(t)$ 
is the distance from a charged proton at position $\vec{x_{i}}$ 
to $\vec{x}$ where the field is evaluated, $\vec{v_{i}}$ is the velocity of the $i$-th
proton inside the colliding nucleus. $R_{i}$ is the magnitude of $\vec{R_{i}}$.
 The summation runs over all proton $(N_{proton})$
inside the two colliding nuclei. Following Ref.~\cite{Bzdak:2011yy}  we assume that because 
of the large Lorentz factor ($\gamma\sim 100$) the colliding nuclei are 
highly Lorentz contracted along $z$ direction and all the  colliding 
protons have  same velocity $v_{i}^{first} = \left(0,0,v_{z}\right)$ and 
 $v_{i}^{second} = \left(0,0,-v_{z}\right)$. $v_z$ is related to the c.m 
 energy ($\sqrt{s_{\rm NN}}$) through the relationship 
 ${ v }_{ z }=\sqrt { 1-{ \left( \frac { 2{ m }_{ p } }{ \sqrt { { s }_{\rm NN } }  }  \right)  }^{ 2 } } $,
 where $m_p$ is the proton mass. Note that according to 
 Eq. (\ref{eq:Efield_LW}) and (\ref{eq:Bfield_LW})
 the electric and magnetic fields diverge as ${ \vec { R }  }_{ i }\rightarrow 0$, to 
 remove this singularity we assume a lower value $R_{cut}=$ 0.3 fm as used
 in Ref~\cite{Bzdak:2011yy}. This particular value of $R_{cut}=$ 0.3 fm 
 was fixed as an effective distance between partons and it was found that
 the calculated electromagnetic field has weak dependence for 
$ 0.3fm\le { R }_{ cut }\le 0.6fm$.
We note that the quantities $e\vec{B}$ and $e\vec{E}$ has dimension  $\rm GeV^{2}$ and    
the conversion from $\rm GeV^2$ to Gauss is given by $1 \rm GeV^{2}$ = $5.12 × 10^{19}$ 
Gauss.  

It is customary to use Milne co-ordinate ($\tau=\sqrt{\left(t^2-z^2\right)},x,y,\eta=\frac{1}{2}
ln\left(\frac{t+z}{t-z}\right)$) in heavy ion collisions. For our case we shall concentrate on the 
mid-rapidity region (z$\approx$0) where $t\sim\tau$.

By using the MC-Glauber model we also compute the 
fluid energy density in transverse plane 
from the position of wounded nucleons. 
This is a common practice to initialise 
energy density for e-by-e hydrodynamics simulations. Since the positions of
 the wounded nucleons~($N_{\rm wound}$) are like delta function in co-ordinate space, 
 in order to calculate the energy density profile for hydrodynamics simulations one
  need to smooth the initial profile by introducing  Gaussian smearing for every
  colliding nucleons. 
The fluid energy density $\varepsilon$  is parameterised as 
\begin{equation}
\varepsilon \left( x,y,\sigma_g ,\vec{b} \right) =k\sum _{ i=1 }^{ { N }_{wound} }{ { e }^{ -\frac { { \left( x-{ x }_{ i }(\vec{b}) \right)  }^{ 2 }+{ \left( y-{ y }_{ i }(\vec{b}) \right)  }^{ 2 } }{ 2{ \sigma_g  }^{ 2 } }  } } ,
\label{Eq:IniEnergyHydro}
\end{equation}
here $x,y$ is the co-ordinate of computational grid,
$x_{i}(\vec{b}),y_{i}(\vec{b})$ are the co-ordinate of wounded nucleons
for an impact parameter $\vec{b}$, 
$\sigma_g$ is the Gaussian smearing which is taken to be 0.5 fm (unless stated otherwise) 
for our calculation. $k$ is a constant which is 
tuned to match the initial central energy density for event averaged case. We 
estimate $k$=6 which results in the initial central energy density 
$40 GeV/fm^{3}$ for $b=0$ fm collision. This is the typical value of initial energy 
density used in e-by-e hydrodynamics model to reproduce the experimental measured
charged particle multiplicity in Au-Au collisions at $\sqrt{s_{\rm NN}}$=200 GeV for 
an initial time $\tau_{i}\sim$ 0.5 fm \cite{Roy:2012jb}.  The same 
$k$ factor is used to calculate the initial energy density 
for all other impact parameter. 

Once we calculate the electromagnetic field and the fluid energy density 
in the transverse plane, the plasma 
$\sigma(x,y,\vec{b}) = \frac{B^{2}(x,y,\vec{b})}{2\varepsilon(x,y,\vec{b})}$ 
is readily obtained for each event. For our case we only considered the 
transverse components $B_x$ and $B_y$ to calculate the 
total magnetic energy density, since $B_z\ll B_x,B_y$.
 As mentioned, the hydrodynamics expansion of the QGP fluid starts after a time $\sim$ 0.5 fm, 
and because of the relativistic velocities of the charged protons the produced
magnetic fields decays very quickly. If there is no conducting medium then 
the magnetic field decay as $\sim t^{-3}$. But in presence of a conducting medium
the decay can be substantially delayed \cite{Tuchin:2013apa}. 
However, the thermodynamic and transport properties of the 
nuclear matter right after the collision upto the time when the system  
reaches local thermal equilibrium is  poorly known. 
Thus we investigate in our study two different scenarios 
when calculating  $\sigma(x,y,\vec{b})$. From now on we will
omit $\vec{b}$ in the expression of $\sigma$, and because of 
spherical symmetry of the colliding nuclei we omit the vector 
arrow and simply write $b$ for the impact parameter. 

(i) In the first scenario, following Ref.~\cite{Tuchin:2013apa} we assume that the matter 
in pre-equilibrium phase  has finite electrical conductivity and the field 
components are evaluated at $\tau_{i}=$0.5 fm  by reducing the magnitudes of the 
initial magnetic field (at $\tau$=0 fm) to  0.1 times. 
This is a simplification of the actual scenario, 
since the time evolution of the fields depend on the electrical conductivity, 
the impact parameter and on the Lorentz gamma($\gamma$) of the collisions. 
According to Ref.~\cite{Tuchin:2013apa} the initial 
electromagnetic field produced in a b=7 fm collision and for an electrical 
conductivity  $\sigma_{el}$=5.8 MeV reduce $\sim$50$\%$
to its original value after $\tau\sim$0.5 fm.  Note that for simplification 
in the numerical simulation we have ignored the impact parameter 
dependence of the evolution of electromagnetic field in medium as was
discussed in Ref.~\cite{Tuchin:2013apa}. 

(ii) In the second scenario, we assume the magnetic field is evolved in vacuum (zero 
electrical conductivity) until the hydrodynamics expansion starts. For this
case we reduced the magnitude of the initial electromagnetic field 
0.01 times.

We note that in reality the  situation may lie in between the 
above mentioned two scenarios. From now on we denote the 
first and second scenario by medium and vacuum respectively.

We consider 1000 nucleus-nucleus 
collisions for our present calculation for each impact parameter.

\section{Results and discussion}

\begin{figure}
\includegraphics[scale=0.55]{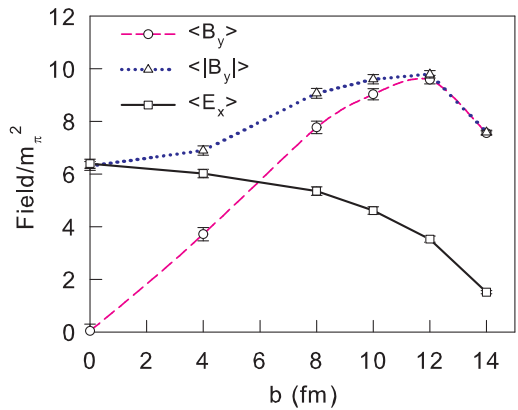}
\caption{Impact parameter dependence of event averaged magnetic
 and electric fields at the centre of the fireball for Au-Au collisions at 
 $\sqrt{s_{\rm NN}}$=200 GeV.}
\label{fig:EventAvMagX0Y0}
\end{figure}

At first we shall concentrate on the electromagnetic fields 
computed at the  centre of the fireball (i.e. at point $\rm x=y=0$ in 
our computational grid). Fig.~\ref{fig:EventAvMagX0Y0}
shows the event averaged value of magnetic 
and electric fields as a function of impact parameter $b$. The $B_y$ , 
its absolute value $|B_y|$, and $x$ component of the electric field $E_x$ are shown 
by pink dashed, blue dotted, and black solid lines respectively.  We note that 
our result is consistent with the result of Ref.~\cite{Bzdak:2011yy}. 
We also checked other components of electric and magnetic fields 
and they are found to be consistent with Ref.~\cite{Bzdak:2011yy}.

\begin{figure}
\includegraphics[scale=0.55]{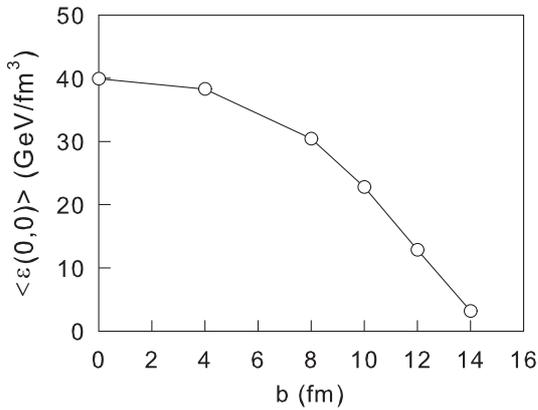}
\caption{Impact parameter dependence of event averaged central energy density 
($<\varepsilon\left(0,0\right)>$) of fluid for Au-Au collisions at $\sqrt{s_{\rm NN} }$
=200 GeV.}
\label{fig:EventAvEps}
\end{figure}

The electric and magnetic fields are created in high energy
 heavy-ion collisions in presence of the electrically 
 charged protons inside the two colliding nucleus. 
Whereas both neutron and protons inside the colliding nuclei  
deposit energy in the collision zone as a result of elastic and inelastic 
collisions among them.  
Since the positions of protons in the 
colliding nucleus is different with that of the positions 
of all nucleons, resulting spatial distribution of the electromagnetic 
field is expected to differ from that of the initial fluid energy density. 
Fig.~\ref{fig:EventAvEps} shows the event averaged value 
of fluid energy density at point ($\rm x=y=0$) as a function of impact parameter
$b$. The energy density is obtained from Eq.~\ref{Eq:IniEnergyHydro}
for $k=$6. This specific value of $k$ was chosen in order to obtain the 
central energy density $\sim$ 40  $\rm GeV/fm^{3}$ for $b=0$ fm 
collisions. From previous studies \cite{Roy:2012jb} we note that the initial central 
energy density for central ($0-5\%$ centrality which corresponds to $b\sim$ 2 fm)
 Au-Au collisions requires $\varepsilon\sim 40 ~\rm GeV/fm^{3}$ at initial time 
 $\tau_{i} = 0.6 ~\rm fm$ at the centre of the fireball ($\rm x=y=0$)
to reproduce the experimentally measured charged hadron multiplicity 
at $\sqrt{s_{\rm NN}}=200 \rm GeV$. However, we note that a different 
initial time ($\tau_{i}$) will give different initial energy density \cite{Shen:2010uy},
 in that case the magnitude of magnetic field at $\tau_{i}$ will also be different. 
From Fig.~\ref{fig:EventAvMagX0Y0} and Fig.~\ref{fig:EventAvEps} we notice that
fluid energy density decreases whereas the intensity of magnetic field 
increases with $b$. It is thus expected that $\sigma(x,y,b)$ will reach its
maximum value for $b\sim$ 12 fm.  So far we have shown the event average 
$\vec{E}(x,y), \vec{B}(x,y)$ and  $\varepsilon\left(x,y\right)$ at the center of the collision zone.

\begin{figure}
\includegraphics[scale=2.1]{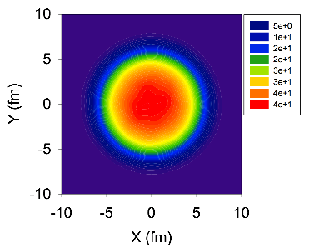}
\includegraphics[scale=2.1]{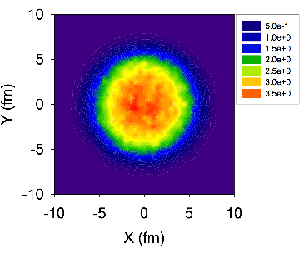}
\includegraphics[scale=2.1]{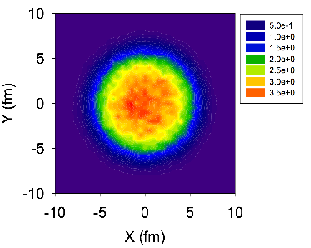}
\caption{Top Panel: 1000 event averaged initial energy density of QGP
 for b=12 fm Au-Au collisions at $\sqrt{s_{\rm NN}}$=200 GeV.
Middle Panel: 1000 event averaged magnetic field energy density calculated from 
y component of the magnetic field
for b=12 fm Au-Au collisions at $\sqrt{s_{\rm NN}}$ GeV.
Bottom Panel: Same as middle panel but for the $x$ component of the magnetic fields $B_x$.  }
\label{fig:EventAvb0}
\end{figure}

\begin{figure}
\includegraphics[scale=2.1]{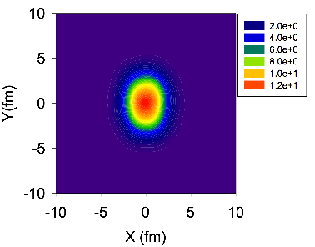}
\includegraphics[scale=2.1]{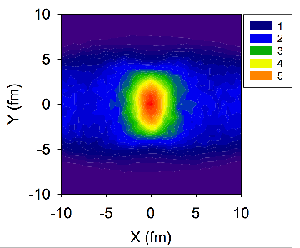}
\includegraphics[scale=2.1]{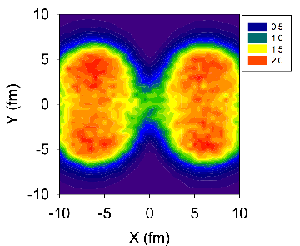}
\caption{Top Panel: 1000 event averaged initial energy density of QGP
 for Au-Au at $\sqrt{s_{\rm NN}}$
=200 GeV for impact parameter b=12 fm collisions.
Middle Panel:  event averaged magnetic field energy density calculated from 
y component of the magnetic field
for b=12 fm Au-Au collisions at $\sqrt{s_{\rm NN}}$ GeV.
Bottom Panel: Same as middle panel but for the $x$ component of the magnetic fields $B_x$.  }
\label{fig:EventAvb12}
\end{figure}

Top panel of Fig.~\ref{fig:EventAvb0}  shows the event averaged 
$\varepsilon(x,y)$ for $\rm b=$0 fm collisions. Since the Au nucleus 
is almost spherical in shape, a head on Au-Au collision  deposits 
energy in an almost circular zone. Different colour schemes in the 
legend denotes the energy density in unit of $\rm GeV/fm^{3}$.
Middle and bottom panel of Fig.~\ref{fig:EventAvb0} shows the 
corresponding magnetic field 
energy density $\frac{B^{2}}{2}$ due to the $y$ and $x$ component of
$\vec{B}$ respectively, where $\vec{B}$ is calculated at time $\tau=0$. We  
observe that the distribution of magnetic field energy is similar to 
the fluid energy density obtained from elastic and inelastic nucleon-nucleon 
collisions in MC-Glauber model. The 
magnetic field energy density due to $B_x$ and $B_y$ is 
peaked at the centre and has a  SO(2) rotational symmetry for $\rm b$=0 fm
collision. This is not surprising since the positions of the protons 
for  $\rm b=$0 fm collisions have such rotational symmetry  about the
centre of the fireball in the
transverse plane for event averaged case. 
The situation for a non-zero impact parameter collisions becomes different .
The overlap zone between the two nuclei becomes elliptical,
 as can be seen from the top panel of Fig.~\ref{fig:EventAvb12} which
corresponds to $\varepsilon(x,y)$ for $\rm b=$12 fm. The middle and bottom 
panel of  Fig.~\ref{fig:EventAvb12} shows the corresponding
energy density for $B_y$ and $B_x$ components. We find that the field energy 
density due to $B_y$ has similar shape as fluid energy density,
 but that due to $B_x$ has maximum in a dumbbell shaped region 
 which is different from the initial fluid energy density.
 
So far we have shown event averaged value of $\varepsilon$ and 
components of $\vec{B}$.
It is not clear from the above discussion whether the magnetic field energy 
density is negligible compared to the initial fluid energy density for every events 
because both $\varepsilon$ and $\frac{B^2}{2}$ are lumpy in the transverse plane 
as shown in Fig.~ \ref{fig:event10}. This leads us to study $\sigma(x,y)$ on e-by-e basis. 
 
 
 \subsection{Event-by-event $\sigma(x,y)$}

\begin{figure}
\includegraphics[scale=0.5]{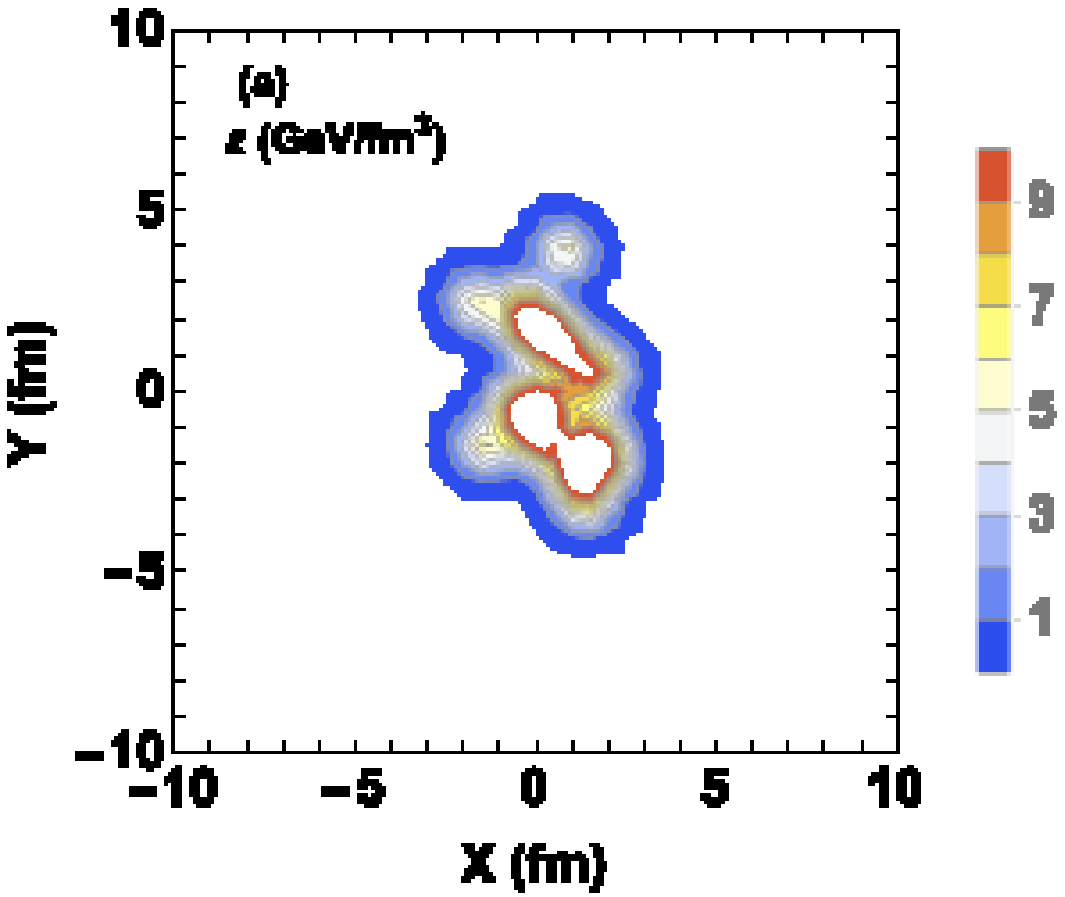}
\includegraphics[scale=0.5]{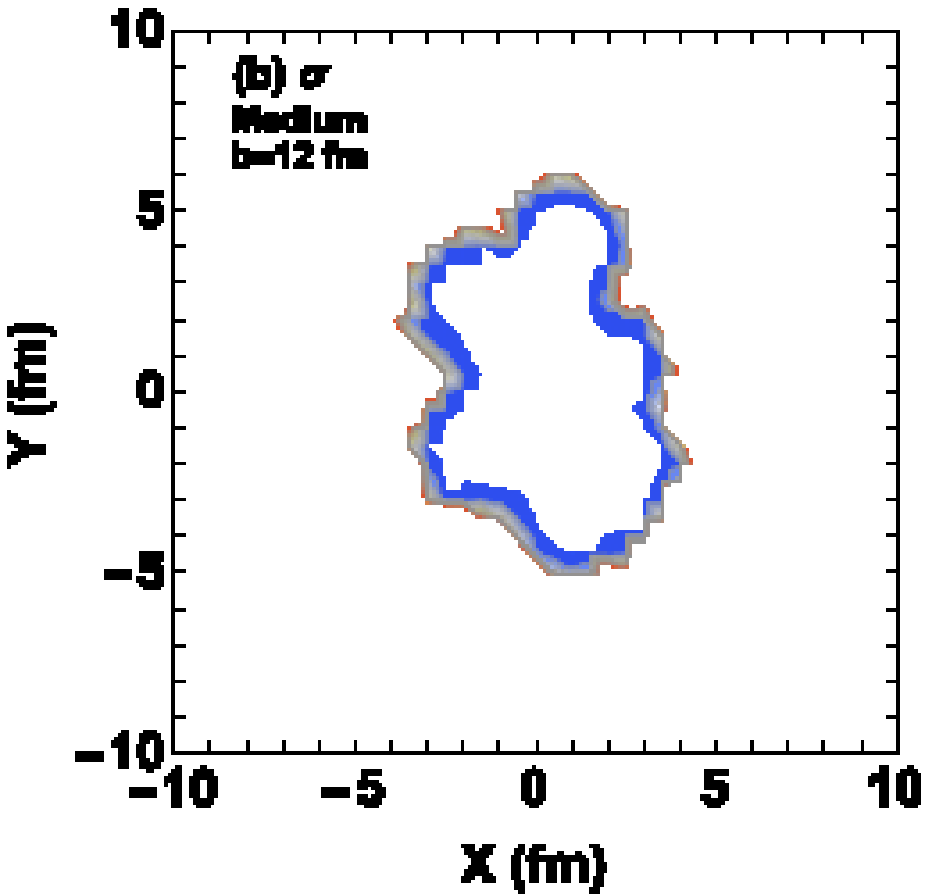}
\includegraphics[scale=0.5]{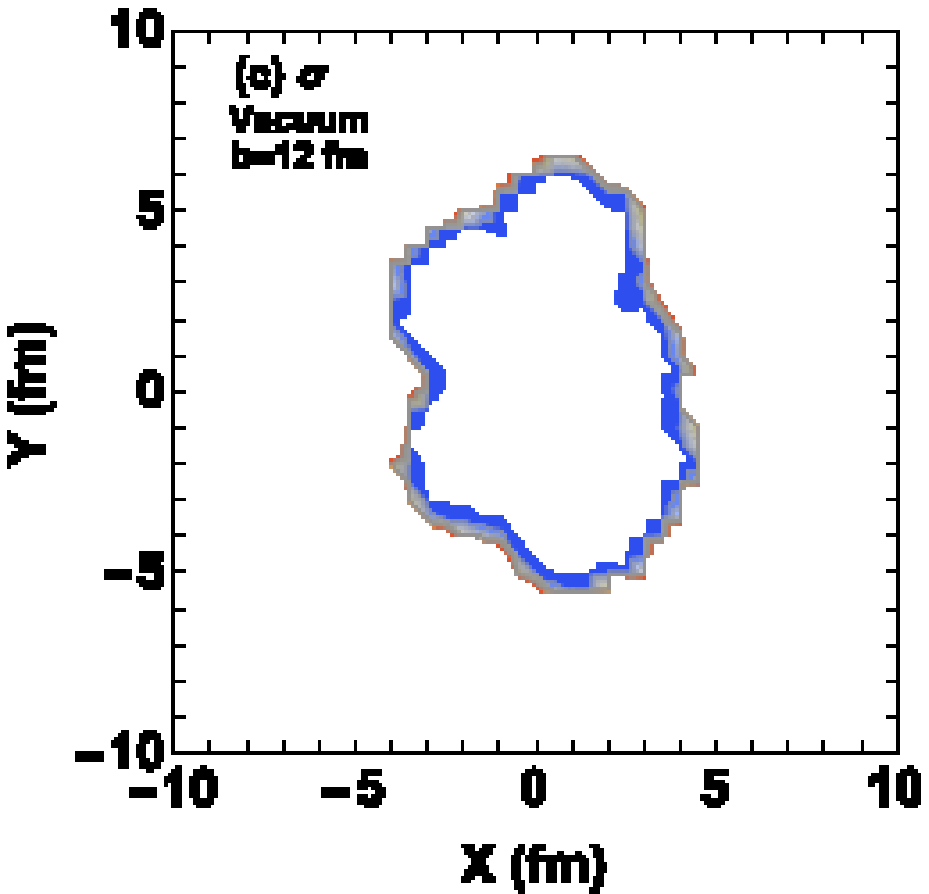}
\caption{ Top panel: fluid energy density , Middle panel: $\sigma\left(x,y\right)$
for the medium , Bottom panel:  $\sigma\left(x,y\right)$ 
 for the vacuum in a single b=12 fm  Au-Au 
 $\sqrt{s_{\rm NN}}$=200 GeV collision. The shaded annular region in middle 
 and bottom panel corresponds to  $0.01\le\sigma\left(x,y\right)\le10$ .}
\label{fig:event10}
\end{figure}

 Top panel of Fig.~\ref{fig:event10}  shows the energy density,
middle and bottom panels show corresponding $\sigma\left(x,y\right)$ 
at $\tau=$0.5 fm for evolution of the magnetic field in medium and 
in vacuum respectively for a single event of b=12 fm collisions. 
The shaded band in the 
  middle and bottom panels correspond to the zones where 
  $0.01\leq\sigma\left(x,y\right)\leq10$ (increasing in the outward direction).
As expected  $\sigma\left(x,y\right)$ reaches its maximum value in regions 
where $\varepsilon\left(x,y\right)$ becomes small. 
 However, note that those regions of large $\sigma\left(x,y\right)$ strongly
 depends on the temporal evolution of the magnetic field from $\tau=0 \rm fm$
 until the hydrodynamics expansion starts at time $\tau_i$. This
 can be seen from the bottom panel of the same figure where the 
 regions of large $\sigma\left(x,y\right)$ moves outward as the magnetic 
 filed for this case decays faster than the case of medium with finite electrical
 conductivity.
\begin{figure}
\includegraphics[scale=0.5]{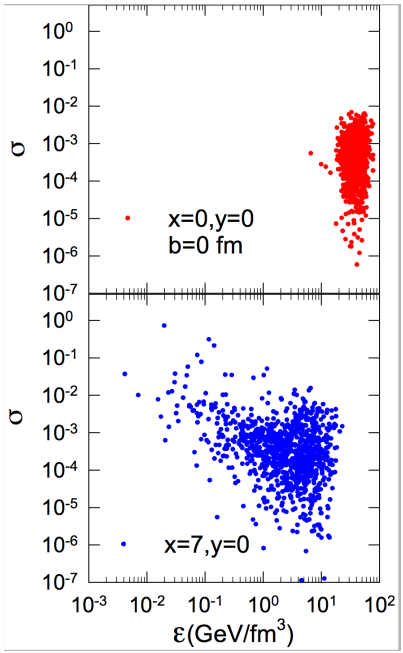}
\caption{Top panel: e-by-e distribution of $\sigma\left(0,0\right)$ as a function 
of $\varepsilon\left(0,0\right)$ for b=0 fm Au-Au collisions at 
$\sqrt{s_{\rm NN}}$ =200 GeV. Bottom panel: same as top panel but for $(\rm x=7,y=0)$.}
\label{fig:b0eventbyeventsigma}
\end{figure}

 We observe here that even if the magnetic field decays quickly (as in  
 vacuum) until the hydrodynamics expansion starts, there is a corona of large
  $\sigma\left(x,y\right)$. It is then important to consider magnetohydrodynamics 
 framework  to investigate further the possible effects of those large $\sigma\left(x,y\right)$ 
 zone on the space-time evolution of the QGP fluid.We expect that since the 
 region of large $\sigma$ seems to lie mostly outside the places where 
 $\varepsilon\left(x,y\right)$ is high there will be minor modification in the 
 transverse evolution of the QGP fluid when the effect of magnetic field is
 taken into account. The above conclusion is made by investigating only
 one particular event, in order to understand the ensemble of events 
 let us look at the e-by-e distribution of $\sigma\left(x,y\right)$
 at the centre ($\rm x=y=0)$ and at the periphery of the collision zone.

Top panel of Fig.~\ref{fig:b0eventbyeventsigma} shows the e-by-e distribution 
of $\sigma\left(0,0\right)$ as a function of $\varepsilon\left(0,0\right)$ for b=0 fm 
collisions. The bottom panel of the same figure shows the event distribution 
of $\sigma\left(7,0\right)$ versus $\varepsilon\left(7,0\right)$.  All results are obtained
for magnetic field evolution in medium.
Naively one expects  that  $\sigma\left(x,y\right)$ and $\varepsilon\left(x,y\right)$ 
should be anti-correlated, i.e., for places where $\varepsilon$ is large $\sigma$ 
will be small and vice-versa, the same conclusion was made
in Ref.~\cite{Voronyuk:2011jd}. But it is clear from Fig.~\ref{fig:b0eventbyeventsigma} 
that there is no such simple relationship between 
$\varepsilon$ and $\sigma$ for b=0 fm collisions in MC-Glauber model . 
In fact, we notice that at the center of the 
collision zone $\varepsilon$ and $\sigma$
are almost uncorrelated. For regions at the periphery of the collision 
zone (bottom panel) we observe similar behaviour, but notice that here
$\sigma\left(7,0\right)$ may reach $\sim$ 1 in some events, whereas 
for $x=y=0$ it never exceed 0.01.

\begin{figure}
\includegraphics[scale=0.5]{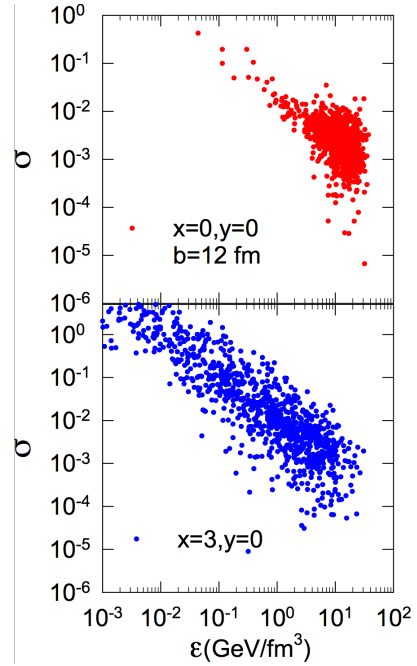}
\caption{Top panel: e-by-e distribution of $\sigma\left(0,0\right)$ as a function 
of $\varepsilon\left(0,0\right)$ for b=12 fm Au-Au collisions at 
$\sqrt{s_{\rm NN}}$ =200 GeV. Bottom panel: same as top panel but for $(\rm x=3,y=0)$.}
\label{fig:b12eventbyeventsigma}
\end{figure}

Now let us discuss the result for $\rm b$=12 fm collisions where the relative 
importance of magnetic field is expected to be highest. Top panel of 
Fig.~\ref{fig:b12eventbyeventsigma} shows the e-by-e distribution of 
$\sigma\left(0,0\right)$ as a function of $\varepsilon\left(0,0\right)$ for
 $\rm  b$=12 fm collisions. We notice that like $\sigma\left(0,0\right)$ 
 distribution for b=0 fm collisions,
most of the events have $\sigma\left(0,0\right)\lesssim $ 0.01.  However, 
for few events $\sigma\left(0,0\right) \sim$ 1. Like $\rm b=0 fm$ collisions here 
we also notice no clear correlation between $\epsilon$ and $\sigma$.
The bottom panel of Fig.~(\ref{fig:b12eventbyeventsigma}) shows the distribution of 
$\sigma\left(3,0\right)$ as a function of $\varepsilon\left(3,0\right)$.
We notice that a considerable number of events have $\sigma \sim$ 1 for this 
case. 
 
\begin{figure}
\includegraphics[scale=0.5]{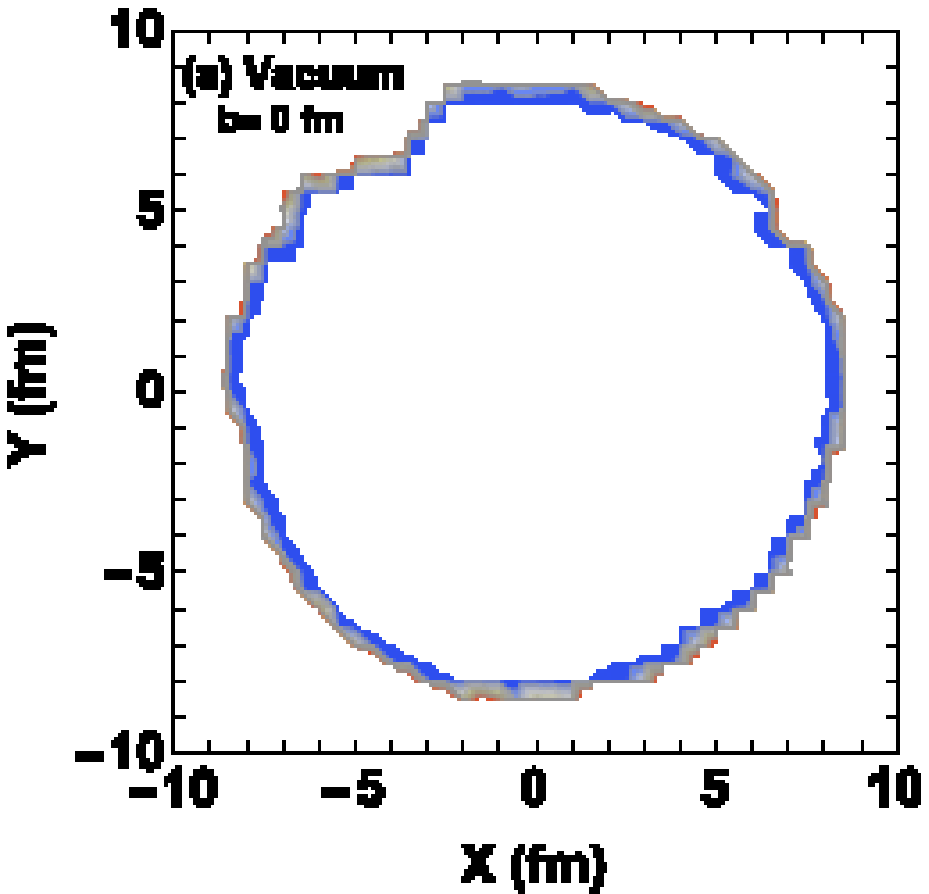}
\includegraphics[scale=0.5]{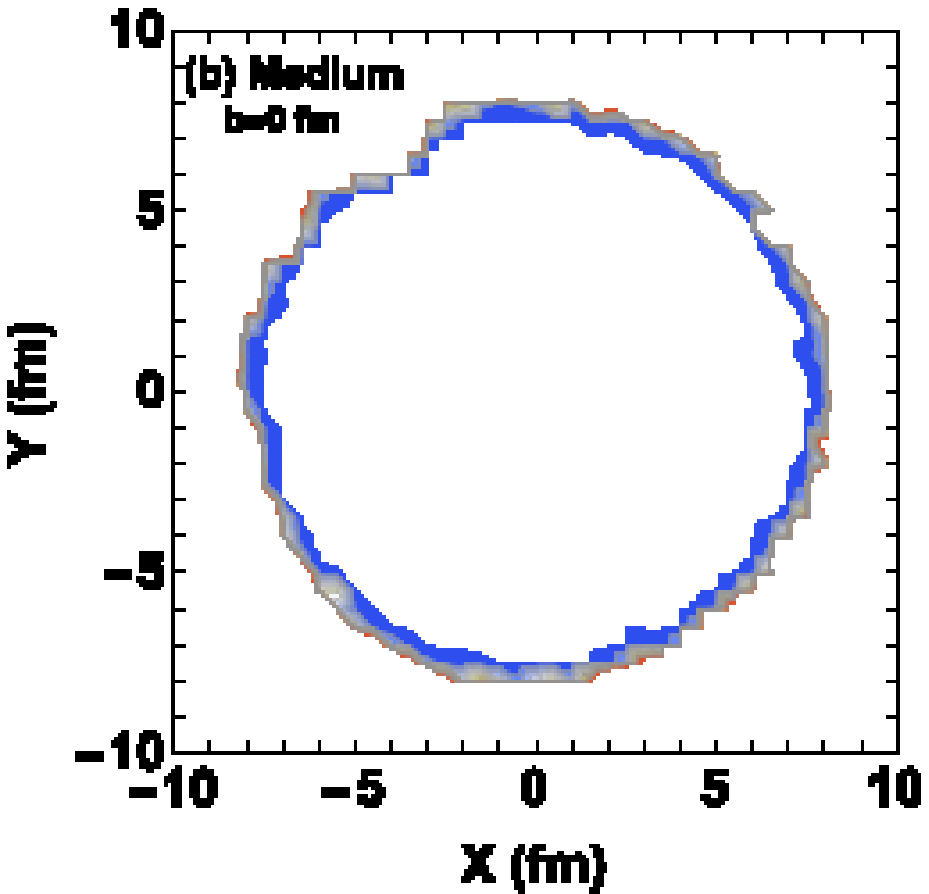}
\caption{Event averaged $\sigma\left(x,y\right)$ in the range 
$0.01\le \sigma \le 10 $ (shaded region)
  for Au-Au collisions of b=0 fm at $\sqrt{s_{\rm NN}}$ = 200 GeV .
  Top panel: for vacuum. Bottom panel: for medium.}
\label{fig:EventAvSigmab0}
\end{figure}

\begin{figure}
\includegraphics[scale=0.5]{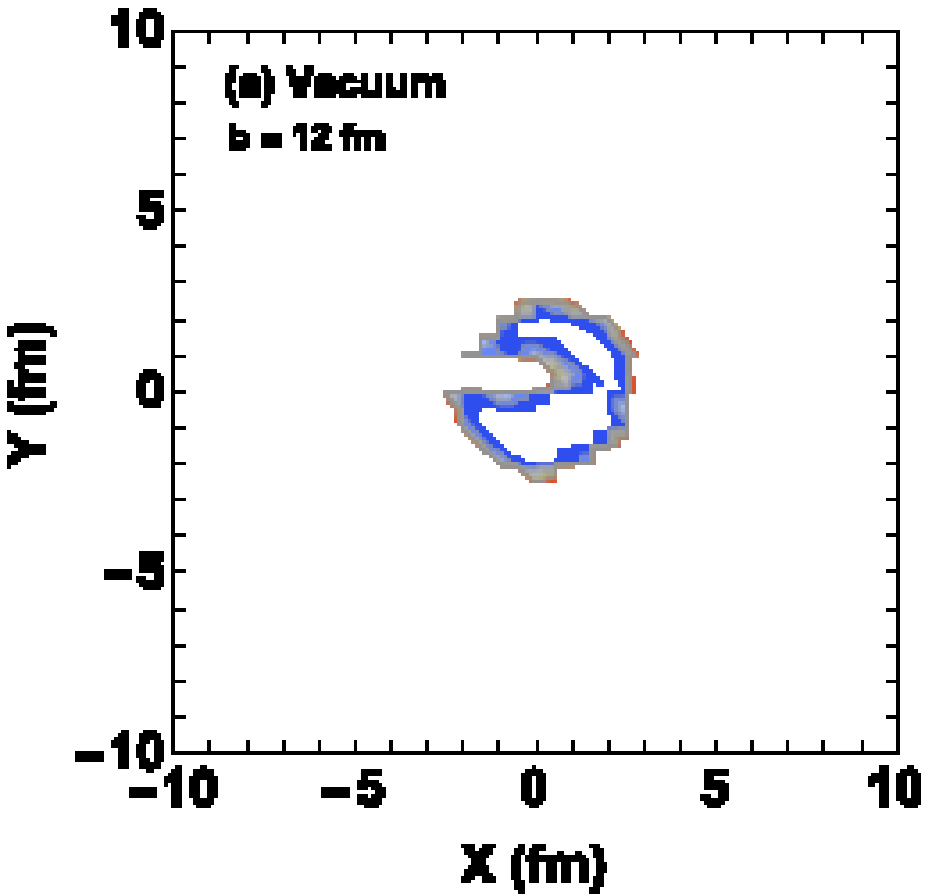}
\includegraphics[scale=0.5]{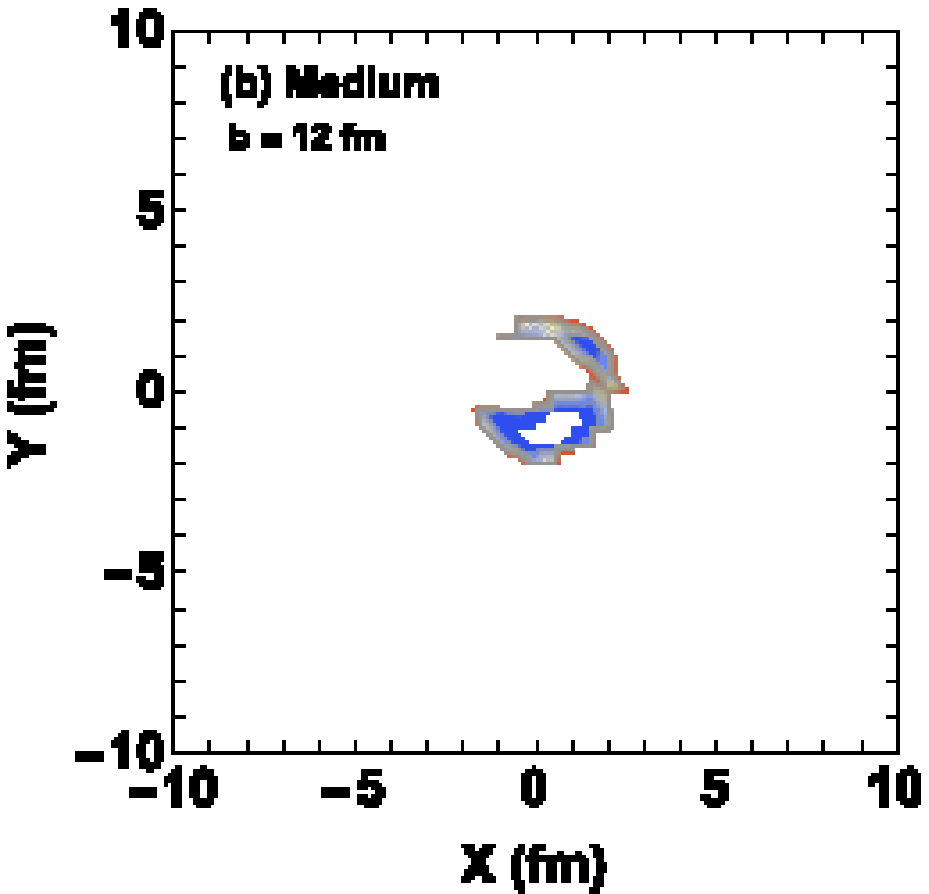}
\caption{Event averaged $\sigma\left(x,y\right)$ in the range 
$0.01\le \sigma \le 10 $ (shaded region)
  for Au-Au collisions of b=12 fm at $\sqrt{s_{\rm NN}}$ = 200 GeV.
  Top panel: for vacuum. Bottom panel: for medium.}
\label{fig:EventAvSigmab12}
\end{figure}

Next we discuss the event averaged transverse profile of $\sigma\left(x,y\right)$ 
for $\rm b$=0 and 12 fm as depicted in Fig.~\ref{fig:EventAvSigmab0} and 
\ref{fig:EventAvSigmab12} respectively.  As expected, the event averaged 
$\sigma\left(x,y\right)$ in the range $0.01\le \sigma \le 10 $  
for b=0 fm collisions (Fig.~\ref{fig:EventAvSigmab0}) form an annular region 
enclosing the high energy density zone of the QGP fluid. 
Top panel of Fig.(\ref{fig:EventAvSigmab0}) shows the result for magnetic field 
evolution in vacuum and the bottom panel shows in medium.
However, $\sigma\left(x,y\right)$  for b=12 fm collisions shows different 
spatial distribution as depicted in Fig.~\ref{fig:EventAvSigmab12}.  The 
non-trivial contour in this case results from the fact that for some events
$\sigma\left(x,y\right)$ becomes very large and the event averaged 
value is dominated by those large $\sigma$. The top and bottom panel
shows the result for vacuum and medium respectively.

 \subsection{Sensitivity of $\sigma(x,y)$ on Gaussian smearing}

\begin{figure}
\includegraphics[scale=0.5]{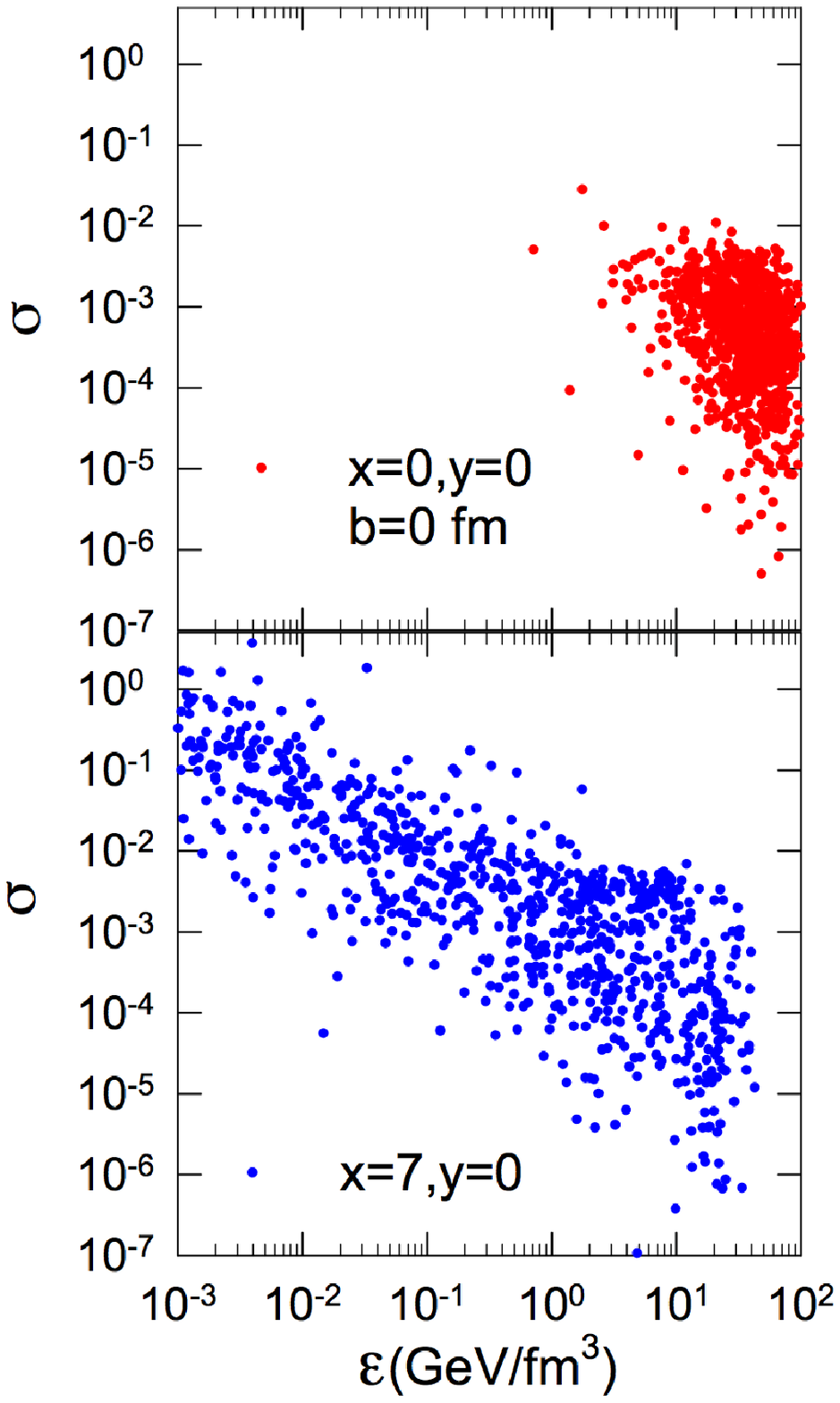}
\caption{Top panel: e-by-e distribution of $\sigma\left(0,0\right)$ as a function 
of $\varepsilon\left(0,0\right)$ for Au-Au  b=0 fm collisions at 
$\sqrt{s_{\rm NN}}$ =200 GeV. Bottom panel: same as top panel but for $(x=7,y=0)$.
$\sigma_g$=0.25 for both the cases.}
\label{fig:sigmag0p25_eps_sigma_b0x0y0}
\end{figure}

\begin{figure}
\includegraphics[scale=0.5]{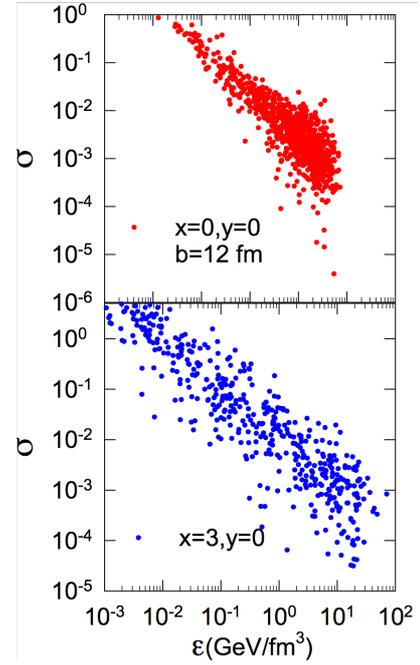}
\caption{Top panel: e-by-e distribution of $\sigma\left(0,0\right)$ as a function 
of $\varepsilon\left(0,0\right)$ for Au-Au  b=12 fm collisions at 
$\sqrt{s_{\rm NN}}$ =200 GeV. Bottom panel: same as top panel but for $(x=3,y=0)$.
$\sigma_g$=0.25 for both the cases.}
\label{fig:sigmag0p25_eps_sigma_b12x0y0}
\end{figure}

The Gaussian smearing  $\sigma_g$ in Eq.~(\ref{Eq:IniEnergyHydro}) 
is a free parameter which is usually taken in the range $\sim$ 0.1-1.0 fm. 
Here we discuss the sensitivity  of our result on Gaussian smearing by 
setting $\sigma_g= \rm 0.25 fm$ which is  taken from a recent 
study~\cite{Chaudhuri:2015twa}. Reducing $\sigma_g$ results in much
lumpy initial energy density hence we expect a different spatial
dependence of $\sigma(x,y)$ compared to the previous case where $\sigma_g
= 0.5 \rm fm$ is used. For $\sigma_g=0.25 \rm fm$ we adjusted $k$ to a 
new value $k$=17 to keep the event-averaged initial central energy density for b=0 fm 
collisions same as before i.e., $\sim 40 \rm GeV/fm^{3}$ . Top panel of Fig.~\ref{fig:sigmag0p25_eps_sigma_b0x0y0} shows the 
e-by-e distribution of $\sigma\left(0,0\right)$ as a function of $\varepsilon\left(0,0\right)$ 
for b=0 fm Au-Au collisions.  Bottom panel shows the same but for $\rm x=7,y=0$.  
Comparing Fig.~\ref{fig:b0eventbyeventsigma} and 
 \ref{fig:sigmag0p25_eps_sigma_b0x0y0} we found 
 that changing $\sigma_g$  from 0.5 fm  to 0.25 fm  
 changes the e-by-e distribution of $\sigma$ vs $\varepsilon$. 
 Since the energy density is more lumpy for  $\sigma_g=$ 0.25 fm 
 than 0.5 fm, the number of events with large $\sigma$ increases.
 To see the effect of changed $\sigma_g$ in peripheral collisions, 
we show the e-by-e distribution of $\sigma$ vs $\varepsilon$ for b=12 fm 
 in Fig.~\ref{fig:sigmag0p25_eps_sigma_b12x0y0}. 
 Top panel of Fig. \ref{fig:sigmag0p25_eps_sigma_b12x0y0} shows the 
 e-by-e distribution of $\sigma\left(0,0 \right) $ vs $\varepsilon\left( 0,0 \right)$
 and the bottom panel shows  e-by-e distribution of 
 $\sigma\left(3,0 \right)$ vs $\varepsilon\left( 3,0\right)$. 
 It is clear that for b=12 fm collisions the correlation between  
 $\sigma$ and $\varepsilon$  at the center ($x=y=0$)
is sensitive to $\sigma_g$, and the maximum value 
of $\sigma\sim$ 1, in contrary to what was observed for the case 
b=0 fm collisions. 

\section{Summary}
We have studied the relative importance of magnetic field energy on initial 
fluid energy density of the QGP by evaluating $\sigma = \frac{B^{2}}{2\varepsilon}$
for Au-Au collisions at $\sqrt{s_{\rm NN}}$= 200 GeV.
The fluid energy density and electromagnetic fields are computed
by using MC-Glauber model. The electromagnetic field and 
initial fluid energy density are calculated by using following parameters:
the cutoff distance $R_{cut}=0.3 \rm fm $, Gaussian smearing parameter 
$\sigma_g=$0.5, (and 0.25 fm) and the scalar multiplicative factor $k=$6,
(and 17). The initial energy density (at time $\tau_i=$ 0.5 fm) for the fluid is fixed
 to $\sim$ 40 $\rm GeV/fm^{3}$.  The ratio of the magnetic field energy density to the 
fluid energy density $\sigma$ is evaluated in the transverse plane for
two different impact parameters $\rm b=$0 , and 12 fm. We find that 
for most of the events, at the centre of the collision zone 
$\sigma\left(0,0\right)$  $\ll$ 1 for both b=0, and 12 fm collisions. However, 
at the periphery of the collision zone where $\varepsilon$ becomes 
small we observed a region of large $\sigma$. For large impact parameter 
collisions $\sigma$ becomes larger for peripheral 
collisions (large $b$) compared to central (small $b$) collisions as a
result of increase in magnetic field and decrease in fluid energy density. 
We observe that in central collisions (b= 0 fm) at the center of collision 
zone $\sigma \ll $1 for most of the events. However, large $\sigma$ is observed 
in the outer regions of collision zone.  In peripheral collisions $\sigma$ 
becomes quite large at both center and periphery of the collision zone.
From this observation we conclude that initial strong magnetic field 
might contribute to the total initial energy density of the Au-Au collisions 
(or other similar heavy ion collisions like Pb-Pb) significantly. However, the true effect
 of large $\sigma$ (or large magnetic fields) will remain
 unclear unless one performs realistic magneto-hydrodynamics simulation 
with the proper initial conditions, for example see Ref.~\cite{Lyutikov:2011vc,Kennel:1983,Roy:2015kma,Spu:2015kma} for some theoretical estimates.
Note that the result in this paper are obtained for a specific model 
of initial conditions (MC-Glauber model) with few free parameters. 
We have not explored all possible allowed 
values of these free parameters. In future we may incorporate other 
initial conditions and more realistic time evolution of the electromagnetic 
fields in the pre-equilibrium phase (as described in Ref.~\cite{Tuchin:2013apa})
 to study the effect of magnetic fields on 
initial fluid energy density distribution. It is also interesting to study the 
similar thing for lower $\sqrt{s_{\rm NN}}$ collisions where the decay of magnetic 
field in vacuum is supposed to be much slower than the present case because 
of the slower speed of colliding nuclei,  and also the corresponding 
initial energy density for such cases is smaller than the present case of Au-Au collisions
at $\sqrt{s_{\rm NN}}=200 \rm GeV$.


\textbf{Acknowledgment:} VR and SP are supported by the Alexander von
Humboldt Foundation. Authors would like to thank Dirk Rischke for 
discussion. 

\end{document}